\documentclass{article}

\usepackage[breaklinks,colorlinks]{hyperref}
\usepackage{multirow}
\usepackage{pgfplots}
\usepackage{amsmath}
\usepackage{amssymb}
\usepackage{enumitem}
\usepackage{fullpage}
\usepackage{comment}
\usepackage{adjustbox}
\usepackage{rotating}
\usepackage[skip=6pt]{caption}
\usepackage{microtype}
\usepackage{mathtools}
\mathtoolsset{centercolon}
\usepackage{url}
\usepackage[numbers,comma,sort&compress]{natbib}
\usepackage{tikz}
\usepackage{soul}

\usepackage{xy}
\xyoption{matrix}
\xyoption{frame}
\xyoption{arrow}
\xyoption{arc}

\usepackage{ifpdf}
\ifpdf
\else
\PackageWarningNoLine{Qcircuit}{Qcircuit is loading in Postscript mode.  The Xy-pic options ps and dvips will be loaded.  If you wish to use other Postscript drivers for Xy-pic, you must modify the code in Qcircuit.tex}
\xyoption{ps}
\xyoption{dvips}
\fi

\entrymodifiers={!C\entrybox}

\newcommand{\ket}[1]{{\left\vert{#1}\right\rangle}}
\newcommand{\qw}[1][-1]{\ar @{-} [0,#1]}
\newcommand{\qwx}[1][-1]{\ar @{-} [#1,0]}

\newcommand{\gate}[1]{*+<.6em>{#1} \POS ="i","i"+UR;"i"+UL **\dir{-};"i"+DL **\dir{-};"i"+DR **\dir{-};"i"+UR **\dir{-},"i" \qw}

\newcommand{\control}{*!<0em,.025em>-=-<.2em>{\bullet}}

\newcommand{\ctrl}[1]{\control \qwx[#1] \qw}

\newcommand{\targ}{*+<.02em,.02em>{\xy ="i","i"-<.39em,0em>;"i"+<.39em,0em> **\dir{-}, "i"-<0em,.39em>;"i"+<0em,.39em> **\dir{-},"i"*\xycircle<.4em>{} \endxy} \qw}
\newcommand{\lstick}[1]{*!R!<.5em,0em>=<0em>{#1}}

\newcommand{\Qcircuit}{\xymatrix @*=<0em>}

\newcommand{\changed}[2]{#2}

\newcommand{\eq}[1]{(\ref{eq:#1})}
\renewcommand{\sec}[1]{\hyperref[sec:#1]{Section~\ref*{sec:#1}}}
\newcommand{\fig}[1]{\hyperref[fig:#1]{Figure~\ref*{fig:#1}}}
\newcommand{\tab}[1]{\hyperref[tab:#1]{Table~\ref*{tab:#1}}}
\newcommand{\routine}[1]{\hyperref[#1]{Routine~\ref*{#1}}}

\newcommand{\Gate}[1]{\textsc{#1}}
\newcommand{\hgate}{\Gate{h}}
\newcommand{\pgate}{\Gate{p}}
\newcommand{\tgate}{\Gate{t}}
\newcommand{\notgate}{\Gate{not}}
\newcommand{\cnotgate}{\Gate{cnot}}
\newcommand{\rzgate}{{\Gate{r}}_{z}}

\newcommand{\GF}{\operatorname{GF}}

\newcommand{\ditto}{%
	\tikz{
		\draw [line width=0.12ex] (-0.2ex,0) -- +(0,0.8ex)
		(0.2ex,0) -- +(0,0.8ex);
		\draw [line width=0.1ex] (-0.6ex,0.4ex) -- +(-0.7em,0)
		(0.6ex,0.4ex) -- +(0.7em,0);
	}%
}

\newlength{\localh}
\newlength{\locald}
\newbox\mybox
\def\mp#1#2{\scalebox{1}{\setbox\mybox\hbox{#2}\localh\ht\mybox\locald\dp\mybox\addtolength{\localh}{-\locald}\raisebox{-#1\localh}{\box\mybox}}}

\newcommand{\urlalt}[2]{\href{#2}{\nolinkurl{#1}}}
\newcommand{\arxiv}[1]{\urlalt{arXiv:#1}{http://arxiv.org/abs/#1}}

\begin{document}

\title{Automated optimization of large quantum \\ circuits with continuous parameters}

\author{Yunseong Nam,$^{1,2,3,}$\footnote{Corresponding author: nam@ionq.co}~ Neil J.\ Ross,$^{1,2,4}$ Yuan Su,$^{1,2,5}$ \\
Andrew M.\ Childs,$^{1,2,5}$ and Dmitri Maslov$^{1,2,6}$  \\[2pt]
\small $^{1}$ Institute for Advanced Computer Studies, University of Maryland, College Park, MD 20742, USA  \\
\small $^{2}$ Joint Center for Quantum Information and Computer Science, University of Maryland, \\[-2pt]
\small College Park, MD 20742, USA \\
\small $^{3}$ IonQ, Inc., College Park, MD 20740, USA \\
\small $^{4}$ Department of Mathematics and Statistics, Dalhousie University, Halifax, NS B3H 4R2, Canada \\
\small $^{5}$ Department of Computer Science, University of Maryland, College Park, MD 20742, USA \\
\small $^{6}$ National Science Foundation, Alexandria, VA 22314, USA}

\date{}

\maketitle


\begin{abstract}
We develop and implement automated methods for optimizing quantum circuits of the size and type expected in quantum computations that outperform classical computers.  We show how to handle continuous gate parameters and report a collection of fast algorithms capable of optimizing large-scale quantum circuits.  For the suite of benchmarks considered, we obtain substantial reductions in gate counts. In particular, we provide better optimization in significantly less time than previous approaches, while making minimal structural changes so as to preserve the basic layout of the underlying quantum algorithms. Our results help bridge the gap between the computations that can be run on existing hardware and those that are expected to outperform classical computers.
\end{abstract}

\section{Introduction}
\label{sec:intro}

Quantum computers have the potential to dramatically outperform classical computers at solving certain problems.  Perhaps their best-known application is to the task of factoring integers: whereas the fastest known classical algorithm is superpolynomial \cite{ar:nfs}, Shor's algorithm solves this problem in polynomial time \cite{ar:shor}, providing an attack on the widely-used RSA cryptosystem.

Even before the discovery of Shor's algorithm, quantum computers were proposed for simulating quantum mechanics \cite{ar:feynman}. By simulating Hamiltonian dynamics, quantum computers can study phenomena in condensed-matter and high-energy physics, quantum chemistry, and materials science.  Useful instances of quantum simulation are likely accessible to smaller-scale quantum computers than classically-hard instances of the factoring problem.

These and other potential applications \cite{www:algorithmzoo} have helped motivate significant efforts toward building a scalable quantum computer.  Two quantum computing technologies, superconducting circuits \cite{www:IBM} and trapped ions \cite{ar:deb}, have matured sufficiently to enable fully programmable universal devices, albeit currently of modest size.  Several groups are actively developing these platforms into larger-scale devices, backed by significant investments from both industry \cite{www:G0,www:I0,Intel,Micro} and government \cite{www:WH,www:UK,www:euro}.  Thus, it is plausible that quantum computations involving tens or even hundreds of qubits will be carried out in the not-too-distant future \cite{www:I,www:G}.

Experimental quantum information processing remains a difficult technical challenge, and the resources available for quantum computation will likely continue to be expensive and severely limited for some time.  To make the most out of the available hardware, it is essential to develop implementations of quantum algorithms that are as efficient as possible.

Quantum algorithms are typically expressed in terms of \emph{quantum circuits}, which describe a computation as a sequence of elementary quantum logic gates acting on qubits.  There are many ways of implementing a given algorithm with an available set of elementary operations, and it is advantageous to find an implementation that uses the fewest resources.  While it is imperative to develop algorithms that are efficient in an abstract sense and to implement them with an eye toward practical efficiency, large-scale quantum circuits are likely to have sufficient complexity to benefit from automated optimization.

In this work, we develop software tools for reducing the size of quantum circuits, aiming to improve their performance as much as possible at a scale where manual gate-level optimization is no longer practical.  Since global optimization of arbitrary quantum circuits is QMA-hard \cite{ar:jwb}, our goal is more modest: we apply a set of carefully chosen heuristics to reduce the gate counts, often resulting in substantial savings.

We apply our optimization techniques to several types of quantum circuits. Our benchmark circuits include components of quantum algorithms for factoring and computing discrete logarithms, such as the quantum Fourier transform, integer adders, and Galois field multipliers. We also consider circuits for the product formula approach to Hamiltonian simulation \cite{ar:lloyd,ar:berryetal}.  In all cases, we focus on circuit sizes likely to be useful in applications that outperform classical computation.  Our techniques can help practitioners understand which implementation of an algorithm is most efficient in a given application.  

While there has been considerable previous work on quantum circuit optimization (as detailed in section ``Comparison with prior approaches''), we are not aware of prior work on automated optimization that has targeted large-scale circuits such as the ones considered here.  Moreover, extrapolation of previously-reported runtimes suggests it is unlikely that existing quantum circuit optimizers would perform well for such large circuits.  We perform direct comparisons by running our software on the same circuits optimized in Ref.~\cite{ar:amm}, showing that our approach typically finds smaller circuits in less time. In addition, to the best of our knowledge, our work is the first to focus on automated optimization of quantum circuits with continuous gate parameters.

\section{Results}
\label{sec:results}

\begin{figure}[t]
\centering
\includegraphics{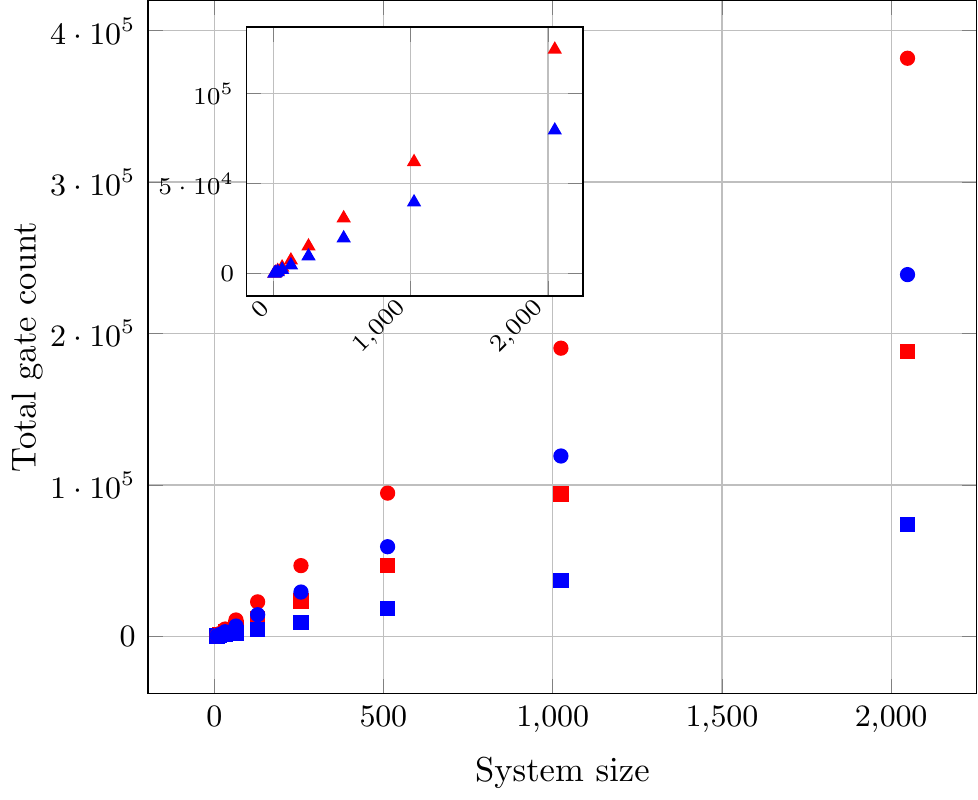}
  \caption{Total gate count for the approximate quantum Fourier transform (QFT, inset), Quipper library adder, and Fourier-based adders (QFA). The points in red/blue represent gate counts before/after optimization and the symbols square/circle/triangle represent gate counts for the Quipper library adder/QFA/QFT, respectively.}
  \label{fig:adders}
  \vspace{-2mm}
\end{figure}

\begin{figure}[t]
  \centering 
  \includegraphics{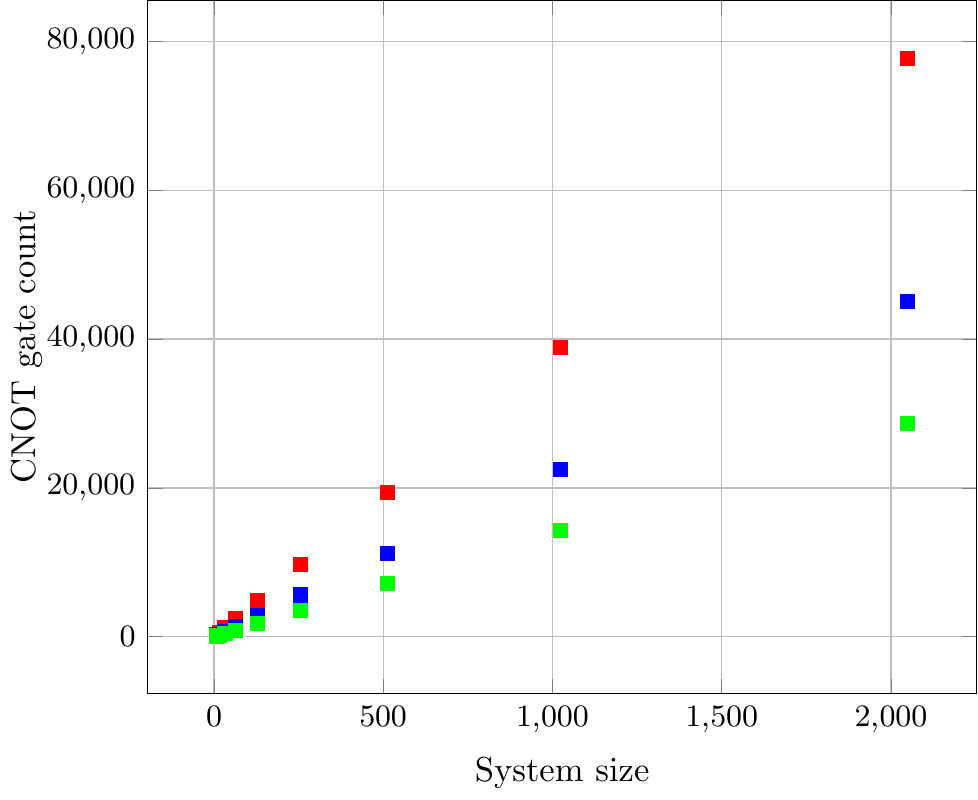}
  \caption{Number of $\cnotgate$ gates for Quipper library adders. The points in red/blue/green represent the gate counts in pre-/post-Light/post-Heavy optimization, respectively.}
  \label{fig:cnot_QA}
  \vspace{-2mm}
\end{figure}

\begin{table}[t]
  \centering
  \caption{Light optimization of adder circuits: QFA (top) and Quipper library (bottom).}
  \label{tab:add}
\adjustbox{center}
{\begin{tabular}[h] {| c | c | c | c | c  | c | c| c|}
 \hline
&  \multicolumn{6}{|c|}{Gate Counts for Approximate QFA}  & \\ \cline{2-7}
& \multicolumn{3}{|c|}{Before Optimization} &\multicolumn{3}{|c|}{After Optimization} & Software Runtime \\ \cline{2-7}
$n$ & $\cnotgate$ & $\rzgate$ & $\hgate$ & $\cnotgate$ & $\rzgate$ & $\hgate$ & (seconds) \\ \hline\hline
8 & 184 & 276 & 16 & 184 & 122 & 16 & $< 0.001$ \\ \hline
16 & 716 & 1,074 & 32 & 716 & 420 & 32 & 0.001 \\ \hline
32 & 1,900 & 2,850 & 64 & 1,900 & 1,076 & 64 & 0.002 \\ \hline
64 & 4,268 & 6,402 & 128 & 4,268 & 2,388 & 128 & 0.004 \\ \hline
128 & 9,004 & 13,506 & 256 & 9,004 & 5,012 & 256 & 0.08 \\ \hline
256 & 18,476 & 27,714 & 512 & 18,476 & 10,260 & 512 & 0.018 \\ \hline
512 & 37,420 & 56,130 & 1024 & 37,420 & 20,756 & 1,024 & 0.045 \\ \hline
1,024 & 75,308 & 112,962 & 2,048 & 75,308 & 41,748 & 2,048 & 0.115 \\ \hline
2,048 & 151,084 & 226,626 & 4,096 & 151,084 & 83,732 & 4,096 & 0.215 \\ \hline
\end{tabular}}
\smallskip
\adjustbox{center}
{\begin{tabular}[h] {| c | c | c | c | c | c | c | c | c| c|}
 \hline
 & \multicolumn{8}{|c|}{Gate Counts for Quipper Library Adder}  & \\ \cline{2-9}
& \multicolumn{4}{|c|}{Before Optimization} &\multicolumn{4}{|c|}{After Optimization} & Software Runtime \\ \cline{2-9}
$n$ & $\cnotgate$ & $\tgate$ & $\hgate$ & $\pgate$ & $\cnotgate$ & $\tgate$ & $\hgate$ & $\pgate$ & (seconds) \\ \hline\hline
8 & 243 & 266 & 76 & 0 & 143 & 56 & 28 & 12 & 0.001 \\ \hline
16 & 547 & 602 & 172 & 0 & 319 & 120 & 60 & 28 & 0.003 \\ \hline
32 & 1,155 & 1,274 & 364 & 0 & 671 & 248 & 124 & 60 & 0.014 \\ \hline
64 & 2,371 & 2,618 & 748 & 0 & 1,375 & 504 & 252 & 124 & 0.057 \\ \hline
128 & 4,803 & 5,306 & 1,516 & 0 & 2,783 & 1,016 & 508 & 252 & 0.244 \\ \hline
256 & 9,667 & 10,682 & 3,052 & 0 & 5,599 & 2,040 & 1,020 & 508 & 1.099 \\ \hline
512 & 19,395 & 21,434 & 6,124 & 0 & 11,231 & 4,088 & 2,044 & 1,020 & 5.292 \\\hline
1,024 & 38,851 & 42,938 & 12,268 & 0 & 22,495 & 8,184 & 4,092 & 2,044 & 25.987 \\ \hline
2,048 & 77,763 & 85,946 & 24,556 & 0 & 45,023 & 16,376 & 8,188 & 4,092 & 145.972 \\ \hline
\end{tabular}}
\end{table}

We implemented our optimizer detailed in section ``Methods'' \footnote{Readers are strongly encouraged to read section ``Methods'' for technical descriptions of the optimization algorithms and their implementation details, used throughout this section.} in the Fortran programming language and tested it using three sets of benchmark circuits. All results were obtained using a machine with a 2.9 GHz Intel Core i5 processor and 8 GB of 1867 MHz DDR3 memory, running OS X El Capitan.

We considered quantum circuits that include components of Shor's integer factoring algorithm, namely the \emph{quantum Fourier transform} (\emph{QFT}) and the integer adders.  We also considered circuits for the \emph{product formula} (\emph{PF}) approach to Hamiltonian simulation \cite{ar:berryetal}. In both cases, we focused on circuit sizes likely to be useful in applications that outperform classical computation, and ran experiments with different types of adders and product formulas.  Finally, we considered a set of benchmark circuits from Ref.~\cite{ar:amm}, consisting of various arithmetic circuits (including a family of Galois field multipliers) and implementations of multiple-control Toffoli gates.  Files containing circuits before and after optimization are available at \cite{github}.

To check correctness of our optimizer, we verified the functional equivalence (i.e., equality of the corresponding unitary matrices) of various test circuits before and after optimization.  Of course, such a test is only feasible for circuits with a small number of qubits.  We performed this test for all $8$-qubit benchmarks in \tab{add} and \tab{heavy}, all $10$-qubit benchmarks in \tab{pf}, and the following benchmarks from \tab{tpar}: Mod $5_4$, VBE-Adder$_3$, CSLA-MUX$_3$, RC-Adder$_6$, Mod-Red$_{21}$, Mod-Mult$_{55}$, Toff-Barenco$_{3 .. 5}$, Toff-NC$_{3 .. 5}$, GF($2^4$)-Mult, and GF($2^5$)-Mult.

\subsection{QFT and adders}
\label{sec:qftadders}

\begin{table}[th]
  \small
  \centering
  \caption{Heavy optimization of Quipper library adder.}
  \label{tab:heavy}
\adjustbox{center}
{\begin{tabular}[h] {| c | c | c | c | c | c | c | c | c| c|}
 \hline
 & \multicolumn{8}{|c|}{Gate Counts for Quipper Library Adder}  & \\ \cline{2-9}
& \multicolumn{4}{|c|}{Before Optimization} &\multicolumn{4}{|c|}{After Optimization (H)} & Software Runtime \\ \cline{2-9}
$n$ & $\cnotgate$ & $\tgate$ & $\hgate$ & $\pgate$ & $\cnotgate$ & $\tgate$ & $\hgate$ & $\pgate$ & (seconds) \\ \hline\hline
8 & 243 & 266 & 76 & 0 & 94 & 56 & 28 & 12 & 0.006 \\ \hline
16 & 547 & 602 & 172 & 0 & 206 & 120 & 60 & 28 & 0.018 \\ \hline
32 & 1,155 & 1,274 & 364 & 0 & 430 & 248 & 124 & 60 & 0.066 \\ \hline
64 & 2,371 & 2,618 & 748 & 0 & 878 & 504 & 252 & 124 & 0.598 \\ \hline
128 & 4,803 & 5,306 & 1,516 & 0 & 1,774 & 1,016 & 508 & 252 & 4.697 \\ \hline
256 & 9,667 & 10,682 & 3,052 & 0 & 3,566 & 2,040 & 1,020 & 508 & 34.431 \\ \hline
512 & 19,395 & 21,434 & 6,124 & 0 & 7,150 & 4,088 & 2,044 & 1,020 & 307.141 \\\hline
1,024 & 38,851 & 42,938 & 12,268 & 0 & 14,318 & 8,184 & 4,092 & 2,044 & 2,446.336  \\ \hline
2,048 & 77,763 & 85,946 & 24,556 & 0 & 28,654 & 16,376 & 8,188 & 4,092 & 23,886.841 \\ \hline
\end{tabular}}
\end{table}

The QFT is a fundamental subroutine in quantum computation, appearing in many quantum algorithms with exponential speedup.  The standard circuit for the exact $n$-qubit QFT uses $\rzgate$ gates, some with angles that are exponentially small in $n$.  It is well known that one can perform a highly accurate approximate QFT by omitting gates with very small rotation angles \cite{ar:coppersmith}. We choose to omit rotations by angles at most $\pi/2^{13}$ since evidence suggests that using this approximate QFT in the factoring algorithm gives an approach with small failure probability for instances of the sizes we consider \cite{ar:nb}. These small rotations are removed before optimization, so their omission does not contribute to the improvements we report.

In \fig{adders} (inset) we plot total gate counts for the approximate QFT before and after optimization. We observe a savings ratio of larger than 36\% for the QFT with 512 or more qubits. The optimization comes entirely from reducing the number of $\rzgate$ gates, the most expensive resource in a fault-tolerant implementation.

We consider two types of integer adders: an in-place modulo $2^q$ adder as implemented in the Quipper library \cite{ar:quipper} and an in-place adder based on the QFT \cite{ar:draper} (hereafter denoted \emph{QFA}). The QFA circuits use an approximate QFT in which the rotations by angles less than $\pi/2^{13}$ are removed, as described above. Adders are a basic component of Shor's quantum algorithm for integer factoring \cite{bk:nc}.  We report gate counts before and after optimization for the Quipper adders and the QFAs for circuits acting on $2^L$ qubits, with $L$ ranging from 4 to 11.  Adders with $L=10$ are used in Shor's algorithm for factoring 1,024-bit numbers.  Recall that the related RSA-$1024$ challenge remains unsolved \cite{www:RSA}.

The results of Light optimization (see section ``General-purpose optimization algorithms'' for its definition) of the adder circuits are shown in \tab{add} and \fig{adders}. For the Quipper library adders, we used the standard Light optimizer. For the QFA optimization, we instead used a modified Light optimizer with the sequence of routines (see section ``Optimization subroutines'' for detail) $\ref{Rhad},\ref{Rdouble},\ref{Rsingle},\ref{Rdouble},\ref{Rhad},\ref{Rsingle}$, omitting the final three routines $\ref{Rlight},\ref{Rdouble},\ref{Rsingle}$ of the full Light optimizer.  We did this because we saw no additional gate savings from those routines in small instances ($n\leq256$).

Observe that the simplified Quipper library adder outperforms the QFA by a wide margin, suggesting that it may be preferred in practice. For the Quipper library adder, we see a reduction in the $\tgate$ gate count by a factor of up to $5.2$. We emphasize that this reduction is obtained entirely by automated means, without using any prior knowledge of the circuit structure. Since Shor's integer factoring algorithm is dominated by the cost of modular exponentiation, which in turn relies primarily on integer addition, this optimization reduces the cost of executing the overall factoring algorithm implemented using the Quipper library adder by a factor of more than 5.

We also applied the Heavy optimizer (see section ``General-purpose optimization algorithms'' for its definition) to the QFT and adder circuits. For the QFT and QFA circuits, the Heavy setting does not improve the gate counts.  The results of the Heavy optimization for the Quipper adder are shown in \tab{heavy}.  We find a reduction in the $\cnotgate$ count by a factor of $2.7$, compared to a factor of only $1.7$ for the Light optimization.  \fig{cnot_QA} illustrates the total $\cnotgate$ counts of the Quipper library adder before optimization, after Light optimization, and after Heavy optimization, showing the reduction in the $\cnotgate$ count by the two types of optimization.

\begin{figure}[t]
  \centering
  \includegraphics{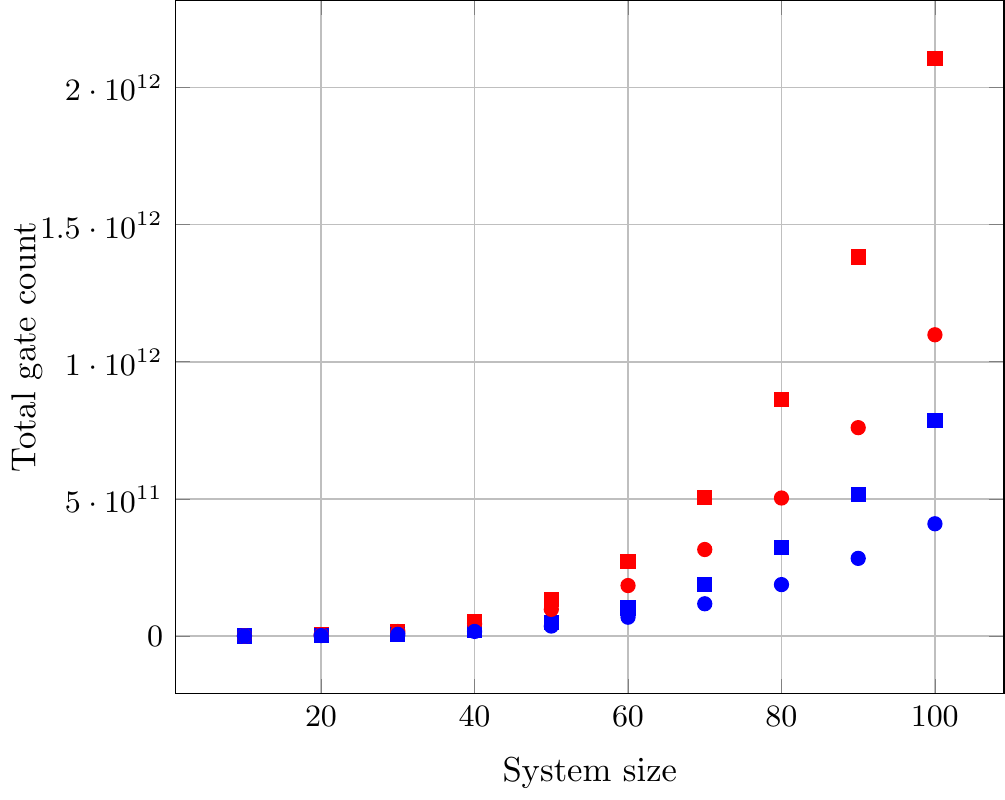}
  \caption{Total gate count for product formula algorithms. The points
    in red/blue represent gate counts before/after optimization and
    the symbols square/circle represent gate counts for the
    \mbox{2nd-/}4th-order formula, respectively.}
  \label{fig:pf}
\end{figure}

\subsection{Quantum simulation}

\begin{table*}[t]
  \small
  \centering
  \caption{Optimization of product formula algorithms, showing the $\cnotgate$ gate count reduction (top) and the $\rzgate$ gate count reduction (bottom). Software runtimes range from 0.004 s (1st-order, $n=10$) to 0.137 s (6th-order, $n=100$). The Clifford gate reduction ranges from 62.5\% for Hadamard and 75\% for Phase gates (for the 1st-order formula, independent of $n$) to 75\% for Hadamard and 85\% for Phase gates (for the 6th-order formula, again independent of $n$).  The notation ``($\times$ 1000)'' indicates that the gate counts for the 1st-order formula are in units of thousands (no rounding errors). The notation ``(L)'' denotes the standard Light optimization.}
  \label{tab:pf}
\adjustbox{center}
{\tabcolsep=0.015cm
{\scriptsize
\begin{tabular}[t]{|c|c|c|c|c|c|c|c|c|}
 \hline
 &\multicolumn{8}{|c|}{$\cnotgate$ Counts for Product Formula Algorithms} \\
 \cline{2-9}
 & \multicolumn{2}{|c|}{1st order} & \multicolumn{2}{|c|}{2nd order}
 &\multicolumn{2}{|c|}{4th order} & \multicolumn{2}{|c|}{6th order} \\ \cline{2-9}
 $n$ &   Before ($\!\times\!$1000) & After (L) ($\!\times\!$1000) &  Before & After (L) &  Before & After (L) &  Before & After (L) \\ [0.5ex]  \hline\hline
  10 & 9,600,024 & 9,600,024 & 49,622,280 & 33,081,540 & 82,152,000 & 54,768,020 & 833,073,000 & 555,382,020 \\  \hline
  20 & 307,200,192 & 307,200,192 & 793,571,040 & 529,047,400 & 927,468,000 & 618,312,040 & 8,376,270,000 & 5,584,180,040 \\ \hline
  30 & 2,332,800,648 & 2,332,800,648 & 4,016,805,120 & 2,677,870,140 & 3,830,076,000 & 2,553,384,060 & 32,322,240,000 & 21,548,160,060 \\ \hline
  40 & 9,830,401,536 & 9,830,401,536 & 12,694,063,680 & 8,462,709,200 &  10,477,257,600 & 6,984,838,480 & 84,262,560,000 & 56,175,040,080 \\ \hline
  50 & 30,000,003,000 & 30,000,003,000 & 30,989,866,200 & 20,659,910,900 &  22,869,948,000 & 15,246,632,100 &  177,187,560,000 & 118,125,040,100  \\ \hline
  60 & 74,649,605,184 & 74,649,605,184 & 64,258,513,920 & 42,839,009,400 &  43,278,861,600 & 28,852,574,520 &  325,230,480,000 & 216,820,320,120 \\ \hline
  70 & 161,347,208,232 & 161,347,208,232 & 119,044,086,000 & 79,362,724,140 &  74,215,289,400 & 49,476,859,740 &  543,505,116,000 & 362,336,744,140 \\ \hline
  80 & 314,572,812,288 & 314,572,812,288 & 203,080,443,840 & 135,386,962,720 &  118,409,788,800 & 78,939,859,360 &  847,991,544,000 & 565,327,696,160 \\ \hline
  90 & 566,870,417,496 & 566,870,417,496 & 325,291,230,720 & 216,860,820,660 &  178,795,738,800 & 119,197,159,380 &  1,255,450,374,000 & 836,966,916,180  \\ \hline
  100 & 960,000,024,000 & 960,000,024,000 & 495,789,866,400 & 330,526,577,800 &  258,496,092,000 & 172,330,728,200 &  1,783,355,700,000 & 1,188,903,800,200 \\ \hline
 \multicolumn{9}{c}{} \\
 \hline
 & \multicolumn{8}{|c|}{$\rzgate$ Counts for Product Formula Algorithms} \\
 \cline{2-9}
 & \multicolumn{2}{|c|}{1st order} &\multicolumn{2}{|c|}{2nd order}
 & \multicolumn{2}{|c|}{4th order} &\multicolumn{2}{|c|}{6th order} \\
 \cline{2-9}
  $n$ &  Before ($\!\times\!$1000) & After (L) ($\!\times\!$1000) &  Before & After (L) &   Before & After (L) &   Before & After (L) \\ [0.5ex]  \hline\hline
   10 & 6,400,016 & 6,400,016 &  28,946,330 & 20,675,960 &  47,922,000 & 34,230,010 &  485,959,250 & 347,113,760 \\  \hline
   20 & 204,800,128 & 204,800,128 &  462,916,440 & 330,654,620 &  541,023,000 & 386,445,020 & 4,886,157,500 & 3,490,112,520 \\ \hline
   30 & 1,555,200,432 & 1,555,200,432 &  2,343,136,320 & 1,673,668,830 &  2,234,211,000 & 1,595,865,030 &  18,854,640,000 & 13,467,600,030 \\ \hline
   40 & 6,553,601,024 & 6,553,601,024 &  7,404,870,480 & 5,289,193,240 &  6,111,733,600 & 4,365,524,040 & 49,153,160,000 & 35,109,400,040 \\ \hline
   50 & 20,000,002,000 & 20,000,002,000 &  18,077,421,950 & 12,912,444,300 &  13,340,803,000 & 9,529,145,050 &  103,359,410,000 & 73,828,150,050 \\ \hline
   60 & 49,766,403,456 & 49,766,403,456 &  37,484,133,120 & 26,774,380,860 &  25,246,002,600 & 18,032,859,060 & 189,717,780,000 & 135,512,700,060 \\ \hline
   70 & 107,564,805,488 & 107,564,805,488 &  69,442,383,500 & 49,601,702,570 &  43,292,252,150 & 30,923,037,320 & 317,044,651,000 & 226,460,465,070 \\ \hline
   80 & 209,715,208,192 & 209,715,208,192 &  118,463,592,240 & 84,616,851,680 &  69,072,376,800 & 49,337,412,080 & 494,661,734,000 & 353,329,810,080 \\ \hline
   90 & 377,913,611,664 & 377,913,611,664 &  189,753,217,920 & 135,538,012,890 &  104,297,514,300 & 74,498,224,590 & 732,346,051,500 & 523,104,322,590 \\ \hline
   100 & 640,000,016,000 & 640,000,016,000 &  289,210,755,400 & 206,579,111,100 &  150,789,387,000 & 107,706,705,100 & 1,040,290,825,000 & 743,064,875,100 \\ \hline
\end{tabular}
}
}
\end{table*}

The first explicit polynomial-time quantum algorithm for simulating Hamiltonian dynamics was introduced in \cite{ar:lloyd}.  This approach was later generalized to higher-order product formulas \cite{ar:berryetal}, giving improved asymptotic complexities.  We report gate counts before and after optimization for the PF algorithms of orders 1, 2, 4, and 6 (for orders higher than 1, the order of the standard Suzuki construction is even). For concreteness, we implement these algorithms for a one-dimensional Heisenberg model with periodic boundary conditions in a random, site-dependent magnetic field, evolving the system for the time proportional to its size, and choose the algorithm parameters to ensure the Hamiltonian simulation error is at most $10^{-3}$ using known bounds on the error of the product formula approximation.

The results of Light optimization of product formula algorithms are reported in \tab{pf} and illustrated in \fig{pf}. For these algorithms, we find that Heavy optimization offers no further improvement. The 2nd-, 4th-, and 6th-order algorithms admit a $\sim \!33.3\%$ reduction in the $\cnotgate$ count and a $\sim \!28.5\%$ reduction in the $\rzgate$ count, roughly corresponding to the reductions relevant to physical-level and logical-level implementations. The 1st-order formula algorithm did not exhibit $\cnotgate$ or $\rzgate$ gate optimization. In all product formula algorithms, the number of Phase and Hadamard gates reduced significantly, by a factor of roughly $3$ to $6$.

\subsection{Comparison with prior approaches}
\label{sec:literature}

\begin{table*}
  \small
  \centering
  \begin{minipage}{\textwidth}
  \caption{$\tgate$-par comparison. The names of the algorithms are taken verbatim from Ref.~\cite{ar:amm}, except that we write Toff-Barenco and Toff-NC to denote implementations of multiple-control Toffoli gates from \cite{barenco} and \cite{bk:nc}, respectively.  The notation ``(L)'' denotes the standard Light optimization, whereas ``(H)'' denotes the standard Heavy optimization. The Runtime $t$ is measured in seconds. The symbol \protect\ditto\ indicates that there was no improvement in the Heavy optimization over the Light optimization.  The column marked ``AC'' reports the aggregate cost.  The last column shows the aggregate cost as applied to the best circuit found by the two versions of our algorithm.}
  \label{tab:tpar}
\adjustbox{center}
{{\setlength{\tabcolsep}{.08em}
\scriptsize
\begin{tabular}[h]{|c|c|c|c|c|c|c|c|c|c|c|c|c|c|c|c|}
 \hline
& \multicolumn{3}{|c|}{Pre-Optimization}
&\multicolumn{4}{|c|}{Ref.~\cite{ar:amm} Post-Optimization}
&\multicolumn{4}{|c|}{Our Post-Optimization (L)}
&\multicolumn{4}{|c|}{Our Post-Optimization (H)} \\ \cline{2-16}
Circuit & Total & $\cnotgate$ & $\tgate$ & Total & $\cnotgate$ & $\tgate$ & t(s) & Total & $\cnotgate$ & $\tgate$ & t(s) & Total & $\cnotgate$ & $\tgate$ & t(s) \\ \hline \hline

Mod $5_4$ & 63 & 28 & 28 & 76 & 48 & 16 & $< 0.001$ & 51 & 28 & 16 & $< 0.001$ & \ditto & \ditto & \ditto & 0.001 \\ \hline
VBE-Adder$_{3}$ & 150 & 70 & 70 & 161 & 114 & 24 & 0.001 & 89 & 50 & 24 & $< 0.001$ & \ditto & \ditto & \ditto & 0.001 \\ \hline
CSLA-MUX$_{3}$ & 170 & 80 & 70 & 508 & 425 & 62\footnote{Our simulation found an error in the circuit optimized by $\tgate$-par.  Specifically, the circuit maps $\ket{1024} \mapsto \frac{\ket{1025}+\ket{1030}+\ket{1161}+\ket{1166}}{2}$ whereas it is supposed to perform the mapping $\ket{1024} \mapsto \ket{1088}$.} & 0.001 & 161 & 76 & 64 & $< 0.001$ & 155 & 70 & 64 & 0.009 \\ \hline
CSUM-MUX$_{9}$ & 420 & 168 & 196 & 593 & 411 & 112\footnote{Note that our software reduced the T-count of the original pre-optimization circuit used by $\tgate$-par to 0.  It turned out that the circuit used by $\tgate$-par is incorrect.  In our optimization reported in this table, we used the correct original circuit \cite[Figure 5]{ar:vi}.} & 0.005 & 294 & 168 & 84 & $< 0.001$ & 266 & 140 & 84 & 0.009 \\ \hline
QCLA-Com$_{7}$ & 443 & 186 & 203 & 751 & 583 & 95 & 0.003 & 284 & 132 & 95 & 0.001 & \ditto & \ditto & \ditto & 0.016\\ \hline
QCLA-Mod$_{7}$ & 884 & 382 & 413 & 1,572 & 1,185 & 249 & 0.008 & 636 & 302 & 237 & 0.004 & 624 & 292 & 235 & 0.077 \\ \hline
QCLA-Adder$_{10}$ & 521 & 233 & 238 & 972 & 737 & 162 & 0.018 & 411 & 195 & 162 & 0.002 & 399 & 183 & 162 & 0.044 \\ \hline
Adder$_{8}$ & 900 & 409 & 399 &1,288 & 920 & 215 & 0.004 & 646 & 331 & 215 & 0.004 & 606 & 291 & 215 & 0.101 \\ \hline
RC-Adder$_{6}$ & 200 & 93 & 77 & 326 & 234 & 63 & 0.001 & 142 & 73 & 47 & $< 0.001$ & 140 & 71 & 47 & 0.004 \\ \hline
Mod-Red$_{21}$ & 278 & 105 & 119 & 425 & 301 & 73 & 0.001 & 184 & 81 & 73 & $< 0.001$ & 180 & 77 & 73 & 0.008 \\ \hline
Mod-Mult$_{55}$ & 119 & 48 & 49 & 223 & 166 & 37 & $< 0.001$ & 91 & 40 & 35 & $< 0.001$ & \ditto & \ditto & \ditto & 0.002 \\ \hline \hline
Toff-Barenco$_{3}$ & 58 & 24 & 28 & 82 & 54 & 16 &$< 0.001$ & 42 & 20 & 16 & $< 0.001$ & 40 & 18 & 16 & 0.001 \\ \hline
Toff-NC$_{3}$ & 45 & 18 & 21 & 65 & 41 & 15 &$< 0.001$ & 35 & 14 & 15 & $< 0.001$ & \ditto & \ditto & \ditto & $< 0.001$ \\ \hline
Toff-Barenco$_{4}$ & 114 & 48 & 56 & 141 & 90 & 28 & $< 0.001$ & 78 & 40 & 28 & $< 0.001$ & 72 & 34 & 28 & 0.001 \\ \hline
Toff-NC$_{4}$ & 75 & 30 & 35 & 102 & 63 & 23 & $< 0.001$ & 55 & 22 & 23 & $< 0.001$ & \ditto & \ditto & \ditto & $< 0.001$ \\ \hline
Toff-Barenco$_{5}$ & 170 & 72 & 84 & 206 & 132 & 40 & 0.001 & 114 & 60 & 40 & $< 0.001$ & 104 & 50 & 40 & 0.003 \\ \hline
Toff-NC$_{5}$ & 105 & 42 & 49 & 148 & 94 & 31 & $< 0.001$ & 75 & 30 & 31 & $< 0.001$ & \ditto & \ditto & \ditto & 0.001 \\ \hline
Toff-Barenco$_{10}$ & 450 & 192 & 224 & 517 & 328 & 100 & 0.004 & 294 & 160 & 100 & 0.001 & 264 & 130 & 100 & 0.012 \\ \hline
Toff-NC$_{10}$ & 255 & 102 & 119 & 361 & 232 & 71 & 0.002 & 175 & 70 & 71 & $< 0.001$ & \ditto & \ditto & \ditto & 0.004 \\ \hline \hline
$\GF(2^4)$-Mult & 225 & 99 & 112 & 419 & 324 & 68 & 0.001 & 187 & 99 & 68 & 0.001 & \ditto & \ditto & \ditto & 0.009 \\ \hline
$\GF(2^5)$-Mult & 347 & 154 & 175 & 682 & 535 & 111 & 0.004 & 296 & 154 & 115 & 0.001 & \ditto & \ditto & \ditto & 0.020 \\ \hline
$\GF(2^6)$-Mult & 495 & 221 & 252 & 842 & 649 & 150 & 0.008 & 403 & 221 & 150 & 0.003 & \ditto & \ditto & \ditto & 0.047 \\ \hline
$\GF(2^7)$-Mult & 669 & 300 & 343 & 1,245 & 992 & 217 & 0.031 & 555 & 300 & 217 & 0.004 & \ditto & \ditto & \ditto & 0.105 \\ \hline
$\GF(2^8)$-Mult & 883 & 405 & 448 & 1,560 & 1,256 & 264 & 0.052 & 712 & 405 & 264 & 0.006 & \ditto & \ditto & \ditto & 0.192 \\ \hline
$\GF(2^9)$-Mult & 1,095 & 494 & 567 & 2,096 & 1,701 & 351 & 0.110 & 891 & 494 & 351 & 0.010 & \ditto & \ditto & \ditto & 0.347 \\ \hline
$\GF(2^{10})$-Mult & 1,347 &609 & 700 & 2,655 & 2,176 & 410 & 0.227 & 1,070 & 609 & 410 & 0.009 & \ditto & \ditto & \ditto & 0.429 \\ \hline
$\GF(2^{16})$-Mult & 3,435 & 1,581 & 1,792 & 7,714 & 6,592 & 1,040 & 5.079 & 2,707 & 1,581 & 1,040 & 0.065 & \ditto & \ditto & \ditto & 5.566 \\ \hline
$\GF(2^{32})$-Mult & 13,562 & 6,268 & 7,168 & 37,563 & 33,269 & 4,128 & 602.577 & 10,601 & 6,299 & 4,128 & 1.834 & \ditto & \ditto & \ditto & 275.698 \\ \hline
$\GF(2^{64})$-Mult & 61,629 & 24,765 & 28,672 & 197,674 & 180,892 & 16,448 & 95,447.466 & 41,563 & 24,765 & 16,448 & 58.341 & & & &  \\ \hline
$\GF(2^{128})$-Mult & 246,141 & 98,685 & 114,688 & N/A & N/A & N/A & N/A & 165,051 & 98,685 & 65,664 & 1,744.746 & & & & \\ \hline
$\GF(2^{131})$-Mult & 258,065 & 103,616 & 120,127 & N/A & N/A & N/A & N/A & 173,370 & 103,616 & 69,037 & 1,953.353 & & & & \\ \hline
$\GF(2^{163})$-Mult & 399,021 & 159,900 & 185,983 & N/A & N/A & N/A & N/A & 267,558 & 159,900 & 106,765 & 4,955.927 & & & & \\ \hline
\end{tabular}
}
}
  \end{minipage}
\end{table*}

Quantum circuit optimization is already a well-developed field (see for example \cite{ar:amm,ar:mdmn,ar:pspmh,ar:sm}). However, to the best of our knowledge, no prior work on circuit optimization has considered large-scale quantum circuits of the kind that could outperform classical computers.  For instance, in \cite{ar:amm}, the complexity of optimizing a $g$-gate circuit is $O(g^3)$ (sections 6.1 and 7), making optimization of large-scale circuits unrealistic. Table 3 in \cite{ar:mdmn} shows running times ranging from 0.07 to 1.883 seconds for numbers of qubits from $n=10$ to $35$ and gate counts from $60$ to $368$, whereas our optimizer ran for a comparable time when optimizing the Quipper adders up to $n=256$ with around 23,000 gates, as shown in \tab{add}. Reference~\cite{ar:pspmh} relies on peep-hole optimization using optimal gate libraries. This is expensive, as is evidenced by the runtimes reported in Tables I and II therein, taking already more than $100$ seconds for a 20-qubit, 1,000-gate circuit.

To compare our results with those reported previously, we consider a weighted combination of the $\tgate$ and $\cnotgate$ counts.  While the $\tgate$ gate can be considerably more expensive to implement fault-tolerantly using state distillation \cite{ar:bk}, neglecting the cost of the $\cnotgate$ gates may lead to a significant underestimate.  For a detailed discussion of this issue, see Reference~\cite{newpaper1}.  Specific analyses suggest that a fault-tolerant $\tgate$ gate may be $46$ \cite{ar:fd} to $350$ \cite{ar:oc} times more expensive to implement than a local fault-tolerant $\cnotgate$ gate, with one possible recommendation to regard the cost ratio as $1{:}50$ \cite{e:f}.  The true overhead depends on many details, including the fault tolerance scheme, the error model, the size of the computation, architectural restrictions, the extent to which the implementation of the $\tgate$ gate can be optimized, and whether $\tgate$ state production happens offline so its cost can be (partially) discounted; it is beyond the scope of this paper to account for all these factors.  For a rough comparison, we choose to work with the \emph{aggregate cost} metric defined as follows: $\#\tgate + 0.01\cdot\log{n}\cdot\#\cnotgate$, where $\#\tgate$ is the number of $\tgate$ gates used, $0.01$ accounts for the relative efficiency of the $\cnotgate$ gate with respect to the $\tgate$ gate, $n$ is the number of qubits in the computation, and $\#\cnotgate$ is the number of $\cnotgate$ gates used.  Here the factor of $\log{n}$ underestimates the typical cost of performing gates between qubits in a realistic architecture (whereas the true cost may be closer to $\sqrt[3]{n}$ in three dimensions or $\sqrt{n}$ in two dimensions). Since our approach preserves the structure of the original circuit, this metric should give a conservative comparison with other approaches (such as the $\tgate$-par approach mentioned below) that may introduce long-range gates. Therefore, showing advantage with respect to this aggregate cost can very crudely demonstrate the benefits of our approach to optimization.

We directly compare our results to those reported in \cite{ar:amm}, which aims to reduce the $\tgate$ count and $\tgate$ depth using techniques based on matroid partitioning.  We refer to that approach as \emph{$\tgate$-par}.  We use our algorithms to optimize a set of benchmark circuits appearing in that work and compare the results with the $\tgate$-par optimization, as shown in \tab{tpar}.

The benchmark circuits fall into three categories.  The first set consists of a selection of arithmetic operations.  For these circuits, we obtained better or matching $\tgate$ counts compared to \cite{ar:amm} while also obtaining much better $\cnotgate$ counts.  Note that we excluded the circuit CSLA-MUX$_{3}$ from the comparison since we do not believe $\tgate$-par optimized it correctly (for more detail, see the first footnote in \tab{tpar}). To illustrate the advantage of our approach using the aggregate cost metric, observe that we reduced the cost of the RC-Adder$_6$ circuit from $71.91$ to $49.70$.  The improvement in cost is thus by about $31\%$, mostly due to a reduced $\tgate$ gate count.

The second set of benchmarks consists of multiple-control Toffoli gates. While our optimizer matched the $\tgate$ count obtained by the $\tgate$-par and substantially reduced the $\cnotgate$ count, neither our optimizer nor \cite{ar:amm} could find the best known implementations constructed directly in \cite{newpaper3}.  This is not surprising, given the very different circuit structure employed in \cite{newpaper3}.

The third set of benchmarks contains Galois field multiplier circuits.  \changed{We saw no advantage from the Heavy optimizer over the Light optimizer in the cases we tested, so we did not apply the Heavy optimizer to the four largest instances}{We terminate the Heavy optimizer when its runtime exceeds that of the light optimizer by a factor of 200. Such a timeout occurred when applying our software to the four largest instances of the Galois field multiplier circuits. Because we saw no advantage from the Heavy optimizer over the Light optimizer in the cases we tested, we did not attempt to run the Heavy optimizer on these larger instances any longer} (the corresponding entries are left blank in \tab{tpar}).  Our $\tgate$ count again matches that of the $\tgate$-par optimizer, but our $\cnotgate$ count is much lower, resulting in circuits that are clearly preferred.  For example, the optimized $\GF(2^{64})$ multiplier circuit in \cite{ar:amm} uses 180,892 $\cnotgate$ gates, whereas our optimized implementation uses only 24,765 $\cnotgate$ gates; the aggregate cost is thus reduced from 30,168.59 to 18,326.42 despite no change in the $\tgate$ count, i.e., by about $39\%$.  The reduction comes solely from the $\cnotgate$ gates.  This comparison therefore demonstrates that the discrepancy between simple $\tgate$ count and realistic aggregate cost estimate predicted in theory \cite{newpaper1} is manifested in practice.  The efficiency of our Light optimizer allowed us to optimize of the $\GF(2^{131})$ and $\GF(2^{163})$ multiplier quantum circuits, corresponding to instances of the elliptic curve discrete logarithm problem that remain unsolved \cite{newpaper4}.  Given the reported $\tgate$-par runtimes \cite{ar:amm}, instances of this size appear to be intractable for the $\tgate$-par optimizer.

\changed{A new tool for}{A tool for} $\tgate$ count optimization \cite{ar:hc} was developed shortly after this paper was first made available on the arXiv.  This new result relies on measurement and classical feedback, in contrast to the fully unitary circuits considered in our work.  Moreover, \cite{ar:hc} does not provide $\cnotgate$ counts, making it impossible to give a direct comparison that accounts for both $\tgate$ and $\cnotgate$ gates.  We emphasize that the new work targets $\tgate$ count optimization, \changed{whereas we departed from this simple costing metric due to its known deficiencies}{whereas we departed from this simple costing metric due to the issues documented in} \cite{newpaper1}.  Observe that the optimized $\mathrm{QFT}_4$ circuit of \cite{ar:hc} implements a $4$-qubit QFT transformation using $44$ qubits, suggesting that the $\cnotgate$ gate overhead must be large.  A further significant difference is scalability: while our tool was explicitly developed for and applied to optimize large circuits, \cite{ar:hc} only treats very small circuits---for instance, the largest $\GF$ multiplier optimized in that work is the $7$-bit case, whereas we successfully tackle $\GF$ multipliers with $131$ and $163$ bits, corresponding to unsolved Certicom challenges \cite{newpaper4}.  Another crucial difference is that we use only those qubit-to-qubit interactions already available in the input circuits.  This enables executing optimized circuits in the same architecture as the input circuit, which may be useful for quantum computers over restricted architectures.  In contrast, \cite{ar:hc} introduces new interactions.  Finally, we can handle circuits with arbitrary $\rzgate$ gates, whereas \cite{ar:hc} is limited to Clifford+$\tgate$ circuits.

\subsection{Overall performance}

Our numerical optimization results are summarized across \tab{add}, \tab{heavy}, \tab{pf}, and \tab{tpar}.  These tables contain benchmarks relevant to practical quantum computations that are beyond the reach of classical computers.  In \tab{add} and \tab{heavy} these are the 1,024- and 2,048-qubit QFT and integer adders used in classically-intractable instances of Shor's factoring algorithm \cite{www:RSA}.  In \tab{pf} these include all instances with $n \gtrsim 50$, for which direct classical simulation of quantum dynamics is currently infeasible.  In \tab{tpar} these are Galois field multipliers over binary fields of sizes $131$ and $163$, which are relevant to quantum attacks on unsolved Certicom ECC Challenge problems \cite{newpaper4}.  This illustrates that our optimizer is capable of handling quantum circuits that are sufficiently large to be practically relevant.

Our optimizer can be applied more generally than previous work on circuit optimization. It readily accepts composite gates, such as Toffoli gates (which may have negated controls). It also handles gates with continuous parameters, a useful feature for algorithms that naturally use $\rzgate$ gates, including Hamiltonian simulation and factoring.  Many quantum information processing technologies natively support such gates, including both trapped ions \cite{ar:deb} and superconducting circuits \cite{www:IBM}, so our approach may be useful for optimizing physical-level circuits.

Fault-tolerant quantum computations generally rely on a discrete gate set, such as Clifford+$\tgate$, and optimal Clifford+$\tgate$ implementations of $\rzgate$ gates are already known \cite{ar:kmm, ar:sr}.  Nevertheless, the ability to optimize circuits with continuous parameters is also valuable in the fault-tolerant setting.  This is because optimizing with respect to a natural continuously-parametrized gate set before compiling into a discrete fault-tolerant set will likely result in smaller final circuits.

Finally, unlike previous approaches \cite{ar:amm,ar:hc,ar:mdmn,ar:pspmh}, our optimizer preserves the structure of the original circuit.  In particular, the set of two-qubit interactions used by the optimized circuit is a subset of those used in the original circuit.  This holds because neither the preprocessing step nor our optimizations introduce any new two-qubit gates.  By keeping the types of interactions used under control (in stark contrast to $\tgate$-par, which dramatically increases the set of interactions used), our optimized implementations are better suited for architectures with limited connectivity.  In particular, given a layout of the original quantum circuit on hardware with limited connectivity, this property allows one to use the same layout for the optimized circuit.  We further note that unlike \cite{ar:amm,ar:hc} our optimizer does not increase the number of the $\cnotgate$ gates used.  This can be a crucial practical consideration since a long-range $\cnotgate$ gate can be even more expensive than a $\tgate$ gate, and focusing on $\tgate$ optimization alone may result in circuits whose cost is dominated by $\cnotgate$ gates \cite{newpaper1}.

\section{Discussion}
\label{sec:discussion}

In this paper, we studied the problem of optimizing large-scale quantum circuits, namely those appearing in quantum computations that are beyond the reach of classical computers.  We developed Light and Heavy optimization algorithms and implemented them in software.  Our algorithms are based on a carefully chosen sequence of basic optimizations, yet they achieve substantial reductions in the gate counts, improving over more mathematically sophisticated approaches such as $\tgate$-par optimization \cite{ar:amm}. The simplicity of our approach is reflected in very fast runtimes, especially using the Light version of the optimizer.

We expect that further improvements can lead to even greater circuit optimization, as demonstrated by the Heavy version of our optimizer. To further improve the output, one could revise the routines for reducing $\rzgate$ count by implementing more extensive (and thus more computationally demanding) algorithms for composing stages of $\cnotgate$ and $\rzgate$ gates, possibly with some Hadamard gates included. One may also consider incorporating template-based \cite{ar:mdmn} and peep-hole \cite{ar:pspmh} optimizations.  It may be worthwhile to expand the set of subcircuit rewriting rules and explore the performance of the approach on other benchmark circuits.  Finally, considering the relative cost of different resources (e.g., different types of gates, ancilla qubits) could lead to optimizers that favorably trade off these resources.

\section{Methods}
\label{sec:alg}

In this section, we detail our optimization algorithms and their implementation.  Throughout, we use $g$ to denote the number of gates appearing in a circuit.  We begin by first defining the notations used throughout this section in ``Background''. We then describe in section ``Representations of quantum circuits'' three distinct representations of quantum circuits that we employ.  In section ``Preprocessing'', we describe a preprocessing step used in all versions of our algorithm.  In section ``Optimization subroutines'', we describe several subroutines that form the basic building blocks of our approach.  Section ``General-purpose optimization algorithms'' explains how these subroutines are combined to form our main algorithms.  Finally, in section ``Special-purpose optimizations'', we present two special-purpose optimization techniques that we use to handle particular types of circuits.

\subsection{Background}
\label{sec:background}

A \emph{quantum circuit} is a sequence of quantum gates acting on a collection of qubits.  Quantum circuits are conveniently represented by diagrams in which horizontal wires denote time evolution of qubits, with time propagating from left to right, and boxes (or other symbols joining the wires) represent quantum gates.  For example, the diagram
\begin{equation}
\begin{aligned}
  \scalebox{0.9}{\mbox{\Qcircuit @C=0.5em @R=0.7em {
&\qw &\ctrl{1} &\gate{\rzgate(\theta)} &\qw       &\targ     &\qw                                          &\ctrl{2} &\qw &\qw\\
&\qw &\targ    &\gate{\hgate}                         &\ctrl{1}  &\qw       &\gate{\rzgate(\theta')} &\qw      &\qw &\qw\\
&\qw &\qw      &\qw                                         &\targ     &\ctrl{-2} &\gate{\hgate}                          &\targ    &\qw &\qw }}}
\end{aligned}
\end{equation}
describes a simple three-qubit quantum circuit.

We consider a simple set of elementary gates for quantum circuits consisting of the two-qubit controlled-$\notgate$ gate (abbreviated $\cnotgate$, the leftmost gate in the above circuit), together with the single-qubit $\notgate$ gate, Hadamard gate $\hgate$, and $z$-rotation gate $\rzgate(\theta)$.  Unitary matrices for these gates take the form
\begin{equation}
  \notgate := \begin{pmatrix}0 & 1 \\ 1 & 0\end{pmatrix}, ~~
  \hgate:=\frac{1}{\sqrt{2}}\begin{pmatrix}
    1 & 1\\
    1 & -1
  \end{pmatrix}, ~~
  \rzgate(\theta):=\begin{pmatrix}
    e^{-i\theta/2} & 0 \\
    0 & e^{i\theta/2}
  \end{pmatrix}, ~~ \mbox{and} ~~
  \cnotgate := \begin{pmatrix}
    1 & 0 & 0 & 0\\
    0 & 1 & 0 & 0\\
    0 & 0 & 0 & 1\\
    0 & 0 & 1 & 0\\
  \end{pmatrix},
\end{equation}
where $\theta \in (0, 2\pi]$ is the rotation angle. The phase gate $\pgate$ and the $\tgate$ gate can be obtained from $\rzgate(\theta)$ up to an undetectable global phase as $\rzgate(\pi/2)$ and $\rzgate(\pi/4)$, respectively.  When the rotation angle is irrelevant, we denote a generic $z$-rotation by $\rzgate$.

While we aim to produce quantum circuits over the set of $\notgate$, $\hgate$, $\rzgate$, and $\cnotgate$ gates, we consider input circuits that may also include Toffoli gates.  The Toffoli gate (the top gate in \fig{toffoli}) is described by the mapping $\ket{x,y,z} \mapsto \ket{x,y,z \oplus (x \wedge y)}$ of computational basis states.  We also allow Toffoli gates to have negated controls.  For example, the Toffoli gate with its top control negated (the middle gate in \fig{toffoli}) acts as $\ket{x,y,z} \mapsto \ket{x,y,z \oplus (\bar x \wedge y)}$, and the Toffoli gate with both controls negated (the bottom gate in \fig{toffoli}) acts as $\ket{x,y,z} \mapsto \ket{x,y,z \oplus (\bar x \wedge \bar y)}$.

The cost of performing a given quantum circuit depends on the physical system used to implement it. The cost can also vary significantly between a physical-level (unprotected) implementation and a logical-level (fault-tolerant) implementation. At the physical level, a two-qubit gate is typically more expensive to implement than a single-qubit gate \cite{ar:deb,www:IBM}. We accommodate this by considering the $\cnotgate$ gate count and optimizing the number of the $\cnotgate$ gates in our algorithms.

For logical-level fault-tolerant circuits, the so-called Clifford operations (generated by the Hadamard, Phase, and $\cnotgate$ gates) are often relatively easy to implement, whereas non-Clifford operations incur significant overhead \cite{ar:bk,bk:nc}. Thus we also consider the number of $\rzgate$ gates in our algorithms and try to optimize their count. In fault-tolerant implementations, $\rzgate$ gates are approximated over a discrete gate set, typically consisting of Clifford and $\tgate$ gates.  Optimal algorithms for producing such approximations are known \cite{ar:kmm, ar:sr}. The number of Clifford+$\tgate$ gates required to approximate a generic $\rzgate$ gate depends primarily on the desired accuracy rather than the specific angle of rotation, so it is preferable to optimize a circuit before approximating its $\rzgate$ gates with Clifford+$\tgate$ fault-tolerant circuits.

By minimizing both the $\cnotgate$ and $\rzgate$ counts, we perform optimizations targeting both physical- and logical-level implementations. One might expect a trade-off between these two goals, and in fact we know of instances where such trade-offs do occur. However, in this paper we only consider optimizations aimed at reducing both the $\rzgate$ and $\cnotgate$ counts.

\subsection{Representations of quantum circuits}
\label{sec:representations}

We use the following three representations of quantum circuits:

\begin{itemize}[topsep=2pt,itemsep=-1pt]
\item First, we store a circuit as a list of gates to be applied sequentially (a \emph{netlist}). It is sometimes convenient to specify the circuit in terms of subroutines, which we call \emph{blocks}. Each block can be iterated any number of times and applied to any subset of the qubits present in the circuit.  A representation using blocks can be especially concise since many quantum circuits exhibit a significant amount of repetition. A block is specified as a list of gates and qubit addresses.

We input and output the netlists using both the .qc format of \cite{ar:amm} and the format produced by the quantum programming language Quipper \cite{ar:quipper}.  Both include the ability to handle blocks.

\item Second, we use a \emph{directed acyclic graph} (DAG) representation. The vertices of the DAG are the gates of the circuit and the edges encode their input/output relationships. The DAG representation has the advantage of making adjacency between gates easy to access.

\item Third, we use a generalization of the \emph{phase polynomial} representation of $\{$\cnotgate,\tgate$ \}$ circuits \cite{ar:ammr}. Unlike the netlist and DAG representations, this last representation applies only to circuits consisting entirely of $\notgate$, $\cnotgate$, and $\rzgate$ gates. Such circuits can be concisely expressed as the composition of an affine reversible transformation and a diagonal phase transformation. Let $C$ be a circuit consisting only of $\notgate$ gates, $\cnotgate$ gates, and the gates $\rzgate(\theta_1),\rzgate(\theta_2),\ldots, \rzgate(\theta_\ell)$. Then the action of $C$ on the $n$-qubit basis state $\ket{x_1,x_2,\ldots, x_n}$ has the form
\begin{equation}
    \ket{x_1,x_2,\ldots, x_n} \mapsto e^{ip(x_1,x_2,\ldots,
    x_n)} \ket{h(x_1,x_2,\ldots, x_n)},
\end{equation}
where $h\colon \{0,1\}^n \to \{0,1\}^n$ is an affine reversible function and
\begin{equation}
    \label{eq:star}
    p(x_1,x_2,\ldots,x_n) = \sum_{i=1}^\ell (\theta_i \bmod 2\pi) \cdot f_i(x_1,x_2, \ldots, x_n)
\end{equation}
is a linear combination of affine Boolean functions $f_i\colon \{0,1\}^n \to \{0,1\}$ with the coefficients reduced modulo $2\pi$.  We call $p(x_1,x_2,\ldots,x_n)$ the \emph{phase polynomial} associated with the circuit $C$. For example, the circuit
\begin{equation}
\begin{aligned}
\label{eq:starstar}
\scalebox{1}{\mbox{\Qcircuit @C=0.5em @R=0.7em {
\lstick{x} & \qw & \qw & \ctrl{1} & \qw & \ctrl{1} & \gate{\rzgate(\theta_3)} & \targ & \qw  \\
\lstick{y} & \qw & \gate{\rzgate(\theta_1)} & \targ & \gate{\rzgate(\theta_2)} & \targ & \gate{\rzgate(\theta_4)} & \ctrl{-1} & \qw }}}
\end{aligned}
\end{equation}
can be represented by the mapping
\begin{equation}
    \ket{x,y} \mapsto e^{ip(x,y)} \ket{x\oplus y,y}
\end{equation}
where $p(x,y) = \theta_1 y+ \theta_2(x\oplus y)+\theta_3 x+ \theta_4 y$. (In Ref.~\cite{ar:ammr}, the phase polynomial representation is only considered for $\{\cnotgate, \tgate\}$ circuits, so all $\theta_i$ in the expression \eq{star} are integer multiples of $\pi /4$ and the functions $f_i$ are linear.)
\end{itemize}

We can convert between any two of the above three circuit representations in time linear in the number of gates in the circuit.  Given a netlist, we can build the corresponding DAG gate-by-gate.  Conversely, we can convert a DAG to a netlist by standard topological sorting.  To convert between the netlist and phase polynomial representations of $\{\notgate, \cnotgate, \rzgate\}$ circuits, we use a straightforward generalization of the algorithm of \cite{ar:ammr}.

\subsection{Preprocessing}
\label{sec:preprocessing}

Before running our main optimization procedures, we preprocess the circuit to make it more amenable to further optimization.
Specifically, the preprocessing applies provided the input circuit consists only of $\notgate$, $\cnotgate$, and Toffoli gates
(as is the case for the Quipper adders described in section ``QFT and adders'' and the $\tgate$-par circuit benchmarks described in section ``Comparison with prior approaches'').
In this case, we push the $\notgate$ gates as far to the right as possible by commuting them through the controls of Toffoli gates and the targets of Toffoli and $\cnotgate$ gates.  When pushing a $\notgate$ gate through a Toffoli gate control, we negate that control (or remove the negation if it was initially negated). If this procedure leads to a pair of adjacent $\notgate$ gates, we remove them from the circuit.  If no such cancellation is found, we revert the control negation changes and move the $\notgate$ gate back to its original position.

This $\notgate$ gate propagation leverages two aspects of our optimizer.  First, we accept Toffoli gates that may have negated controls and optimize their decomposition into Clifford+$\tgate$ circuits by exploiting freedom in the choice of $\tgate/\tgate^{\dagger}$ polarities (see section ``Special-purpose optimizations'').  Second, since cancellations of $\notgate$ gates simplify the phase polynomial representation (by making some of the functions $f_i$ in the phase polynomial representation \eq{star} linear instead of merely affine), such cancellations make it more likely that \routine{Rlight} and \routine{Rheavy} in section ``Optimization subroutines'' will find optimizations (since those routines rely on finding matching terms in the phase polynomial representation).

The complexity of this preprocessing step is $O(g)$ since we simply make a single pass through the circuit.

\subsection{Optimization subroutines}
\label{sec:subroutines}

Our optimization algorithms rely on a variety of subroutines that we now describe.  For each of them, we report the worst-case time complexity as a function of the number of gates $g$ in the circuit (for simplicity, we neglect the dependence on the number of qubits and other parameters).  We optimize practical performance by carefully ordering and restricting the subroutines, as we discuss further below.

\begin{figure}[htbp]
\centering
\adjincludegraphics[width=\textwidth,Clip={0.1\width} {.72\height} {0.1\width} {.18\height},scale=1.2]{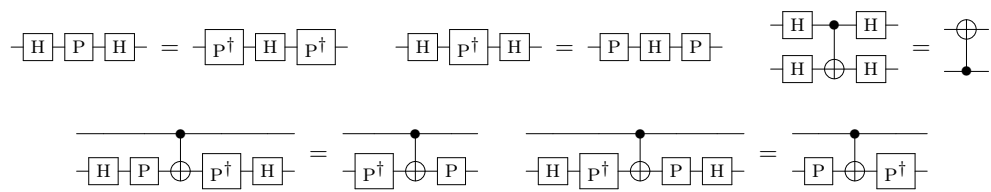}
  \caption{Hadamard gate reductions. The two rules illustrated on the bottom can be applied even if the middle $\cnotgate$ gate is replaced by a circuit with any number of $\cnotgate$ gates, provided they all share the target of the original $\cnotgate$.}
  \label{fig:hadamards}
\end{figure}

\begin{figure}
\centering
\adjincludegraphics[width=\textwidth,Clip={0.1\width} {.70\height} {0.1\width} {.18\height},scale=1.2]{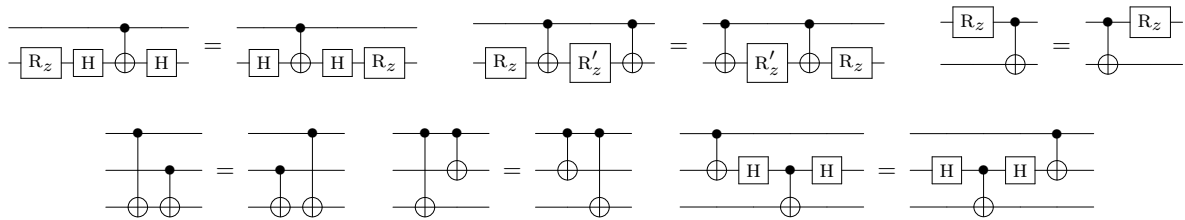}
  \caption{Commutation rules. Top: Commuting an $\rzgate$ gate to the right. Bottom: Commuting a $\cnotgate$ gate to the right.}
  \label{fig:commutations}
\end{figure}

\begin{enumerate}
\item\label{Rhad} \emph{Hadamard gate reduction} \nopagebreak

Hadamard gates do not participate in phase polynomial optimization (\routine{Rlight} and \routine{Rheavy} below) and also tend to hinder gate commutation.  Thus, we use the circuit identities pictured in \fig{hadamards} to reduce the Hadamard gate count.  Each application of these rules reduces the $\hgate$ count by up to 4.  For a given Hadamard gate, we can use the DAG representation to check in constant time whether it is involved in one of these circuit identities.  Thus, we can implement this subroutine with complexity $O(g)$ by making a single pass through all Hadamard gates in the circuit.

\item\label{Rsingle} \emph{Single-qubit gate cancellation} \nopagebreak

Using the DAG representation of a quantum circuit, it is straightforward to determine whether a gate and its inverse are adjacent.  If so, both gates can be removed to reduce the gate count.  More generally, we can cancel two single-qubit gates $U$ and $U^\dagger$ that are separated by a subcircuit $A$ that commutes with $U$.  In general, deciding whether a gate $U$ commutes with a circuit $A$ may be computationally demanding. Instead, we apply a specific set of rules that provide sufficient (but not necessary) conditions for commutation.  This approach is fast and appears to discover many commutations that can be exploited to simplify quantum circuits.

Specifically, for each gate $U$ in the circuit, the optimizer searches for possible cancellations with some instance of $U^\dagger$.  To do this, we repeatedly check whether $U$ commutes through a set of consecutive gates, as evidenced by one of the patterns in \fig{commutations}.  If at some stage we cannot move $U$ to the right by some allowed commutation pattern, then we fail to cancel $U$ with a matched $U^\dag$, so we restore the initial configuration.  Otherwise, we successfully cancel $U$ with some instance of $U^\dagger$.

For each of the $g$ gates $U$, we check whether it commutes through $O(g)$ subsequent positions.  Thus the complexity of the overall gate cancellation rule is $O(g^2)$.  We could make the complexity linear in $g$ by only considering commutations through a constant number of subsequent gates, but we do not find this to be necessary in practice.

We also use a slight variation of this subroutine to merge rotation gates, rather than cancel inverses.  Specifically, two rotations $\rzgate(\theta_1)$ and $\rzgate(\theta_2)$ can be combined into a single rotation $\rzgate(\theta_1 + \theta_2)$ to eliminate one $\rzgate$ gate.

\item\label{Rdouble} \emph{Two-qubit gate cancellation} \nopagebreak

This routine is analogous to \routine{Rsingle}, except that $U$ is a two-qubit gate, which is always $\cnotgate$ in the circuits we consider.  Again its complexity is $O(g^2)$, but may be reduced to $O(g)$ by imposing a maximal size for the subcircuit $A$.

\item\label{Rlight} \emph{Rotation merging using phase polynomials} \nopagebreak

Consider a subcircuit consisting of $\notgate$, $\cnotgate$, and $\rzgate$ gates.  Observe that if two individual terms of its phase polynomial expression satisfy $f_i(x_1,x_2, \ldots , x_n)=f_j(x_1,x_2, \ldots , x_n)$ for some $i \neq j$, then the corresponding rotations $\rzgate(\theta_i)$ and $\rzgate(\theta_j)$ can be merged. For example, in the circuit \eq{starstar}, the first and fourth rotations are both applied to the qubit carrying the value $y$, as evidenced by its phase polynomial representation.  Thus \eq{starstar} goes through the transformation
\begin{equation}\label{eq:ppolymerge}
\begin{aligned}
  \scalebox{1}{\mbox{\Qcircuit @C=0.5em @R=0.7em {
\lstick{x} & \qw & \qw & \ctrl{1} & \qw & \ctrl{1} & \gate{\rzgate(\theta_3)} & \targ & \qw  \\
\lstick{y} & \qw & \gate{\rzgate(\theta_1)} & \targ & \gate{\rzgate(\theta_2)} & \targ & \gate{\rzgate(\theta_4)} & \ctrl{-1} & \qw }}}
  \hspace{2mm}\raisebox{-5mm}{$\mapsto$}\hspace{4mm}
  \scalebox{1}{\mbox{\Qcircuit @C=0.5em @R=0.5em {
\lstick{x} & \ctrl{1} & \qw & \ctrl{1} & \gate{\rzgate(\theta_3)} & \targ & \qw  \\
\lstick{y} & \targ & \gate{\rzgate(\theta_2)} & \targ &
      \gate{\rzgate(\theta_1 + \theta_4)} & \ctrl{-1} & \qw }}}
\end{aligned}
\end{equation}
in which the two rotations are combined. In other words, the phase polynomial representation of circuits reveals when two rotations---in this case, $\rzgate(\theta_1)$ and $\rzgate(\theta_4)$---are applied to the same affine function of the inputs, even if they appear in different parts of the circuit. Then we may combine these rotations into a single rotation, improving the circuit.\footnote{Note that in this particular example, the simplification could have alternatively been obtained using the commutation method described above. However, this is not the case in general.}  We have the flexibility to place the combined rotation at any point in the circuit where the relevant affine function appears. For concreteness, we place it at the first (leftmost) such location.

\end{enumerate}

We next discuss some implementation details for \routine{Rlight}.  To apply this routine, we must identify a subcircuit consisting only of $\{ \notgate, \cnotgate, \rzgate \}$ gates.  We build this subcircuit one qubit at a time, starting from a designated $\cnotgate$ gate.  For the first qubit of this gate, we scan through all preceding and subsequent $\notgate$, $\cnotgate$, and $\rzgate$ gates that act on this qubit, adding them to the subcircuit.  When we encounter a Hadamard gate or the beginning or end of the circuit, we mark a \emph{termination point} and stop exploring in that direction (so that each qubit has one beginning termination point and one ending termination point).  For each $\cnotgate$ gate between this qubit and some qubit that has not yet been encountered, we mark an \emph{anchor point} where the gate acts on the newly-encountered qubit.  We then carry out this process with the second qubit acted on by the initial $\cnotgate$ gate, and repeat the process starting from every anchor point until no new qubits are encountered.

While the resulting subcircuit consists only of $\notgate$, $\cnotgate$, and $\rzgate$ gates, it may not have a phase polynomial representation---specifically, intermediate Hadamard gates on the wires that leave and re-enter the subcircuit can prevent this.  To apply the phase polynomial formalism, we ensure this does not happen using the following pruning procedure.  Starting with the designated initial $\cnotgate$ gate, we successively consider gates both before and after it in the netlist until we encounter a termination point.  Note that we only need to consider $\cnotgate$ gates, since every $\notgate$ and $\rzgate$ gate reached by this process can always be included, as it does not prevent the phase polynomial expression from being applied.  If both the control and target qubits of an encountered $\cnotgate$ gate are within the termination border, we continue.  If the control qubit is outside the termination border but the target qubit is inside, we move the termination point of the target qubit so that the $\cnotgate$ gate being inspected falls outside the border, excluding it and any subsequent gates acting on its target qubit from the subcircuit.  However, when the control is inside the border and the target is outside, we make an exception and do not move the termination point (although we do not include the $\cnotgate$ gate in the subcircuit).  This exception gives a larger $\{ \notgate, \cnotgate, \rzgate \}$ subcircuit that remains amenable to phase polynomial representation.  We illustrate the process of obtaining a suitable subcircuit with the following sample circuit:

\begin{equation}
\begin{aligned}
\label{eq:subcircuitexception}
	\begin{tikzpicture}
	  \node at (0,0) {
	  \mbox{\Qcircuit @C=0.5em @R=0.7em {
			\lstick{q_1} &\qw &\gate{\hgate} &\qw &\targ     &\gate{\rzgate} &\ctrl{1}       &\qw      &\ctrl{1}  &\qw            &\gate{\hgate} &\qw\\
			\lstick{q_2} &\qw &\gate{\hgate} &\gate{\rzgate} &\ctrl{-1} &\ctrl{1}       &\targ         &\ctrl{1} &\targ     &\gate{\rzgate} &\gate{\hgate} &\qw\\
			\lstick{q_3} &\qw &\gate{\hgate} &\gate{\rzgate} &\qw       &\targ          &\gate{\hgate} &\targ    &\qw       &\qw            &\qw           &\qw
			}}
	  };
		\draw[densely dotted] (-2.05,1) -- (-2.05,-1) -- (-.1,-1) -- (-.1,-.3) -- (.6,-.3) -- (.6,.2) -- (.95,.2) -- (.95,-.3) -- (2.23,-.3) -- (2.23,1) -- (-2.05,1);
	\end{tikzpicture}
\end{aligned}
\end{equation}

In the example circuit \eq{subcircuitexception}, suppose we start our search from the first $\cnotgate$ gate acting on the top ($q_1$) and middle ($q_2$) qubits. Traversing $q_1$ to the left, we find an $\hgate$ gate, where we mark a termination point. Traversing $q_1$ to the right, we find two $\cnotgate$ gates, one $\rzgate$ gate, and then an $\hgate$ gate, where we mark a termination point. Observe that neither of the encountered $\cnotgate$ gates joins $q_1$ or $q_2$ to the remaining qubit $q_3$. Next, we repeat the same procedure on $q_2$ from the original $\cnotgate$ gate. To the left we find an $\rzgate$ gate and then an $\hgate$ gate, where we mark a termination point. Traversing to the right, we find a $\cnotgate$ acting on $q_2$ and $q_3$. This $\cnotgate$ reveals additional connectivity, so we mark an anchor point at the target of this $\cnotgate$ gate. Further to the right on the $q_2$ wire, we have three more $\cnotgate$ gates (none of which reveals additional connectivity), an $\rzgate$ gate, and finally an $\hgate$ gate, where we mark a termination point. Next we examine $q_3$. We start from the aforementioned anchor point. To the left, we find an $\hgate$ gate with no further connections to other qubits, where we mark a termination point. To the right, we immediately find an $\hgate$ gate and mark a termination point.

Having built the subcircuit, we go through the netlist representation and prune it.  In this pass, we encounter the fourth $\cnotgate$ gate acting on $q_2$ and $q_3$, where we find that the control is within the border but the target is not. In this case we continue according to the exception handling scheme described in the pruning procedure. This ensures that we include the last $\cnotgate$ gate in the $\{\notgate, \cnotgate, \rzgate\}$ region, while excluding the fourth $\cnotgate$ gate (as indicated by the dotted border in \eq{subcircuitexception}). Thus we discover that the last $\rzgate$ gate appearing in the circuit can be relocated to the very beginning of the circuit on the $q_2$ line, to the right of the leftmost $\hgate$, enabling a phase-polynomial based $\rzgate$ merge (see below for details).

Once a valid $\{\notgate, \cnotgate,\rzgate\}$ subcircuit is identified, we generate its phase polynomial. For each $\rzgate$ gate, we determine the associated affine function its phase is applied to and the location in the circuit where it is applied.  We then sort the list of recorded affine functions.  Finally, we find and merge all $\rzgate$ gate repetitions, placing the merged $\rzgate$ at the first location in the subcircuit that computes the desired affine function.

This procedure considers $O(g)$ subcircuits, and the cost of processing each of these is dominated by sorting, with complexity $O(g\log g)$, giving an overall complexity of $O(g^2 \log g)$ for \routine{Rlight}.  However, in practice the subcircuits are typically smaller when there are more of them to consider, so the true complexity is lower.  In addition, when identifying a $\{\notgate, \cnotgate, \rzgate\}$ subcircuit, we choose to start with a $\cnotgate$ gate that has not yet been included in any of the previously-identified $\{\notgate, \cnotgate, \rzgate\}$ subcircuits, so the number of subcircuits can be much smaller than $g$ in practice. If desired, the overall complexity can be lowered to $O(g)$ by limiting the maximal size of the subcircuit.

As a final step, we reduce all affine functions of phases to linear functions.  This is accomplished using $\notgate$ propagation through $\cnotgate$ and $\rzgate$ gates as follows:
\begin{itemize}
\item $\notgate(a) \cnotgate(a;b) \mapsto \cnotgate(a;b) \notgate(a) \notgate(b)$;
\item $\notgate(b) \cnotgate(a;b) \mapsto \cnotgate(a;b) \notgate(b)$;
\item $\notgate(a) \rzgate(a) \mapsto \rzgate^\dagger(a) \notgate(a)$.
\end{itemize}
Applying this procedure ensures that each affine function $x_{i_1} \oplus x_{i_2} \oplus \cdots \oplus x_{i_k} \oplus 1$ transforms into the corresponding linear function $x_{i_1} \oplus x_{i_2} \oplus \cdots \oplus x_{i_k}$, thereby improving the chance to induce further phase collisions.

We now return to the description of optimization subroutines.

\begin{figure}[t]
\centering
\adjincludegraphics[width=\textwidth,Clip={0.1\width} {.74\height} {0.1\width} {.18\height},scale=1.2]{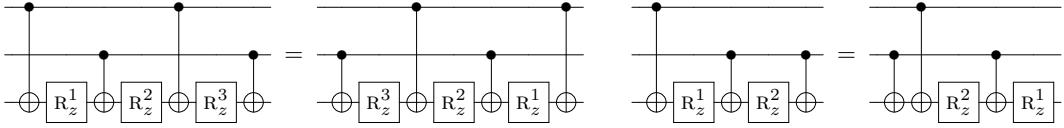}
	\caption{Gate count preserving rewriting rules employed in \routine{Rheavy}.}
	\label{fig:replacement_rules_7}
\end{figure}

\begin{figure}
\centering
\adjincludegraphics[width=\textwidth,Clip={0.1\width} {.28\height} {0.1\width} {.18\height},scale=1.2]{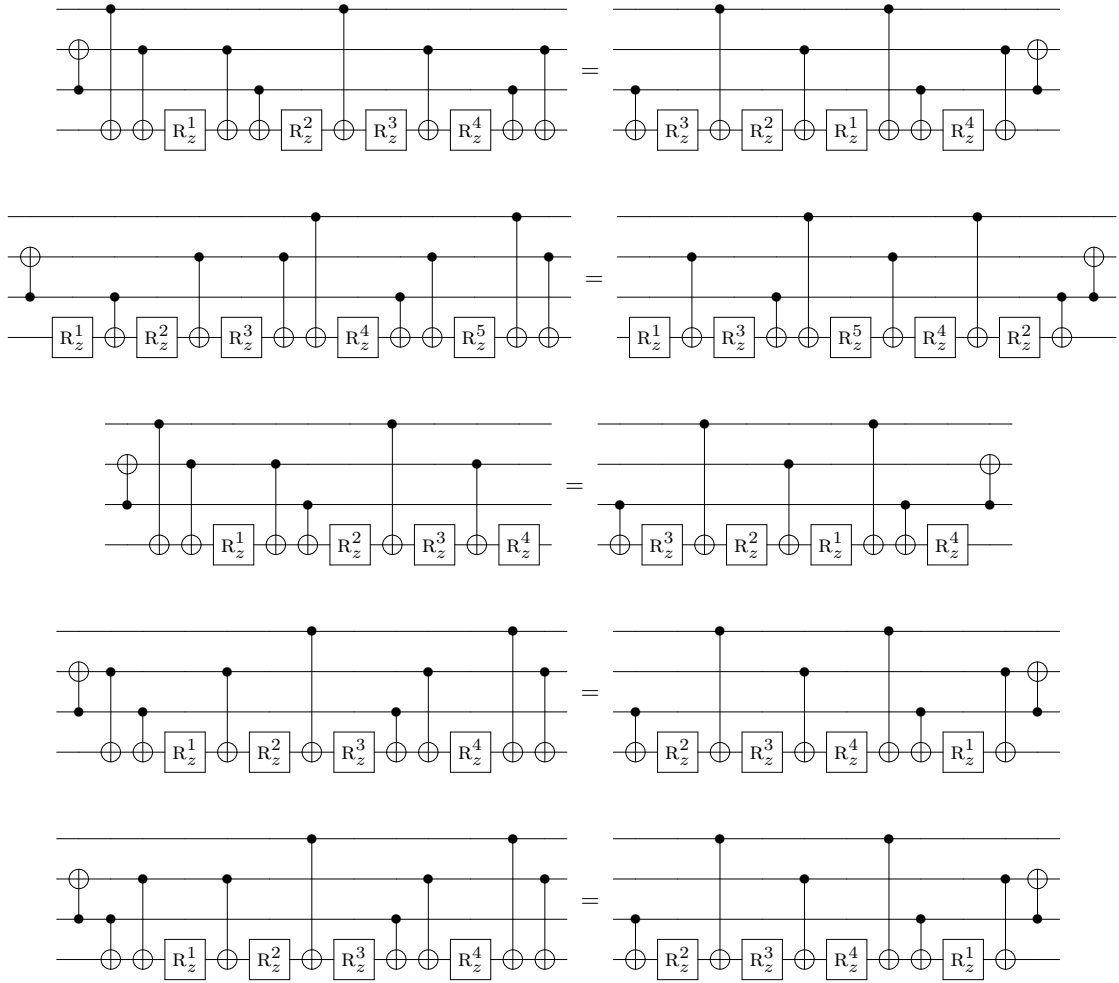}
\caption{Gate count reducing rewriting rules employed in \routine{Rheavy}.}
\label{fig:replacement_rules_8}
\end{figure}

\begin{enumerate}[resume]

\item\label{Rheavy}{\emph{Floating $\rzgate$ gates} } \nopagebreak

In \routine{Rlight}, we keep track of the affine functions associated with $\rzgate$ gates.  More generally, we can record all affine functions that occur in the subcircuit and their respective locations, regardless of the presence of $\rzgate$ gates.  Thus we can identify all possible locations where an $\rzgate$ gate could be placed, not just those locations where $\rzgate$ gates already appear in the circuit. In this ``floating'' $\rzgate$ gate placement picture, we employ three optimization subroutines: two-qubit gate cancellations, gate count preserving rewriting rules, and gate count reducing rewriting rules.

The first of these subroutines is essentially identical to \routine{Rdouble}, except that $\rzgate$ gates are now floatable and we focus on a specific identified subcircuit.  This approach allows us to place $\rzgate$ gates to facilitate cancellations by keeping track of all possible $\rzgate$ gate locations along the way.  In particular, if not placing an $\rzgate$ gate at a particular location will allow two $\cnotgate$ gates to cancel, we simply remove that location from the list of possible locations for the $\rzgate$ gate while ensuring that the reduced list remains non-empty, and perform the $\cnotgate$ cancellation.

We next apply rewriting rules that preserve the gate count (see \fig{replacement_rules_7}) in an attempt to find further optimizations.  While these replacements do not eliminate gates, they modify the circuit in ways that can enable optimizations elsewhere.  The rewriting rules are provided by an external library file, and we identify subcircuits to which they can be applied using the DAG representation.  The replacements are applied only if they lead to a reduction in the two-qubit gate count through one more round of the aforementioned two-qubit cancellation subroutine with floatable $\rzgate$ gates.  Note that the rewriting rules are applicable only with certain floating $\rzgate$ gates at particular locations in a circuit. This subroutine uses floating $\rzgate$ gates to choose those combinations of $\rzgate$ gate locations that lead to reduction in the gate count.

The last subroutine applies rewriting rules that reduce the gate count (see \fig{replacement_rules_8}).  These rules are also provided via an external library file.  Since these rules reduce the gate count on their own, we always perform the rewriting whenever a suitable pattern is found.

The complexity of this three-step routine is upper bounded by $O(g^3)$ since the number of subcircuits is $O(g)$, and within each subcircuit, the two-qubit cancellation (\routine{Rdouble}) has complexity $O(g^2)$. The rewriting rules can be applied with complexity $O(g)$ since, as in \routine{Rhad}, a single pass through the gates in the circuit suffices.  Again, in practice, the number of subcircuits and the subcircuit sizes are typically inversely related, which lowers the observed complexity by about a factor of $g$.  The complexity can also be lowered to $O(g^2)$ by limiting the maximal size of the subcircuit.  The complexity can be further lowered to $O(g \log g)$ by limiting the maximal size of the subcircuit $A$ in the two-qubit gate cancellation (the sorting could still have complexity $O(g\log g)$).

To illustrate how this optimization works, consider the circuit from \eq{ppolymerge} on the right-hand side.  Observe that $\rzgate(\theta_2)$ may be executed on the top qubit at the end of the circuit, leading to the optimization
\begin{equation*}
\begin{aligned}
\scalebox{1}{\mbox{\Qcircuit @C=0.5em @R=0.5em {
\lstick{x} & \ctrl{1} & \qw & \ctrl{1} & \gate{\rzgate(\theta_3)} & \targ & \qw  \\
\lstick{y} & \targ & \gate{\rzgate(\theta_2)} & \targ &
      \gate{\rzgate(\theta_1 + \theta_4)} & \ctrl{-1} & \qw }}}
  \hspace{2mm}\raisebox{-5mm}{$\mapsto$}\hspace{4mm}
  \scalebox{1}{\mbox{\Qcircuit @C=0.5em @R=0.5em {
\lstick{x} & \gate{\rzgate(\theta_3)} & \targ & \qw  &  \gate{\rzgate(\theta_2)} & \qw \\
\lstick{y} & \gate{\rzgate(\theta_1 + \theta_4)} & \ctrl{-1} & \qw & \qw & \qw}}}
\end{aligned}
\end{equation*}
in which the first two $\cnotgate$s cancel.

\end{enumerate}

\subsection{General-purpose optimization algorithms}
\label{sec:algo_gen}

Our optimization algorithms simply apply the subroutines from section ``Optimization subroutines'' in a carefully chosen order.  We consider two versions of the optimizer that we call \emph{Light} and \emph{Heavy}. The Heavy version applies more subroutines, yielding better optimization results at the cost of a higher runtime.  The preprocessing step (see section ``Preprocessing'') is used in both Light and Heavy versions of the optimizer.

The Light version of the optimizer applies the optimization subroutines in the order
\[
    \ref{Rhad},\, \ref{Rdouble},\, \ref{Rsingle},\, \ref{Rdouble},\, \ref{Rhad},\, \ref{Rsingle},\, \ref{Rlight},\, \ref{Rdouble},\, \ref{Rsingle}.
\]
We then repeat this sequence until no further optimization is achieved. We chose this sequence based on the principle that first exposing $\{\cnotgate, \rzgate\}$ gates while reducing Hadamard gates (\ref{Rhad}) allows for greater reduction in the cancellation routines (\ref{Rdouble}, \ref{Rsingle}, \ref{Rdouble}), and in particular frees up two-qubit $\cnotgate$ gates to facilitate single-qubit gate reductions and vice versa.  Applying the replacement rule (\ref{Rhad}) may enable more reductions after the first four optimization subroutines.  We then look for additional single-qubit gate cancellation and merging (\ref{Rsingle}).  This enables faster identification of the $\{\notgate,\cnotgate,\rzgate\}$ subcircuit regions to look for further $\rzgate$ count optimizations (\ref{Rlight}), after which we check for residual cancellations of the gates (\ref{Rdouble}, \ref{Rsingle}).

The Heavy version of the optimizer applies the sequence
\[
    \ref{Rhad},\, \ref{Rdouble},\, \ref{Rsingle},\, \ref{Rdouble},\, \ref{Rhad},\, \ref{Rsingle}, \, \ref{Rheavy}.
\]
Similarly, we repeat this sequence until no further optimization is achieved. The first six steps of the Heavy optimization sequence are identical to that of the Light optimizer.  The difference is that in the Heavy optimizer, we take advantage of floating $\rzgate$ gates. This allows us to find locations for the $\rzgate$ gates that admit better $\cnotgate$ gate reductions, including the use of gate count preserving rewriting rules to expose further gate cancellations and gate count reducing rewriting rules to remove any remaining inefficiency.

We note in passing that the computational overhead incurred due to the circuit representation conversion is minimal.  All conversions can be done in time linear in the circuit size (see section ``Representations of quantum circuits'' for detail).  We keep representations consistent only as necessary.  In \routine{Rhad}--\routine{Rdouble}, we access individual gates using the DAG representation to quickly find reductions.  This allows us to update only the DAG representation to record gate count reductions before continuing with the optimization process.  In \routine{Rlight} and \routine{Rheavy}, we concurrently update both representations on the fly whenever a reduction is found, keeping both the DAG and netlist representations consistent.  This is useful since both routines identify subcircuits that are amenable to reductions using the phase polynomial representation. The identification process requires an up-to-date DAG representation and creating the phase polynomial representation requires an up-to-date netlist representation.  Note that the phase polynomial representation is employed only to aid optimization in the identified subcircuit; it is not necessary to convert the phase polynomial representation back to either the netlist or the DAG representation.  The phase polynomial representation may thus be safely purged when the corresponding subcircuit optimization process is finished.

\subsection{Special-purpose optimizations}
\label{sec:algo_spec}

\begin{figure}[t]
\centering
\adjincludegraphics[width=\textwidth,Clip={0.1\width} {.55\height} {0.1\width} {.18\height},scale=1.2]{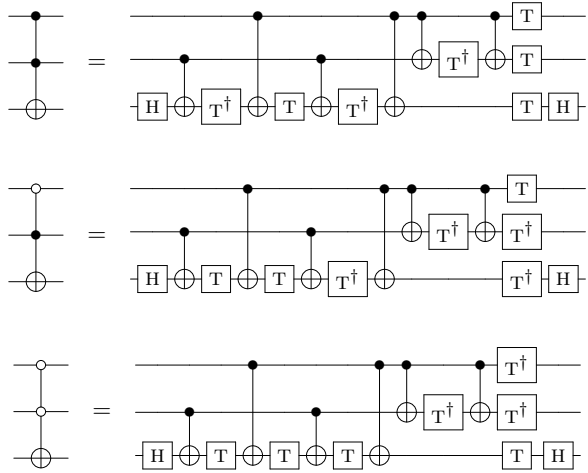}
	\caption{Toffoli gate implementations.}
	\label{fig:toffoli}
\end{figure}

In addition to the general-purpose optimization algorithms described above, we employ two specialized optimizations to improve circuits with particular structures.

\begin{itemize}
\item{\emph{$\mathcal{LCR}$ optimizer:}}
Some quantum algorithms---such as product formula simulation algorithms---involve repeating a fixed block multiple times.  To optimize such a circuit, we first run the optimizer on a single block to obtain its optimized version, $\mathcal{O}$. To find simplifications across multiple blocks, we optimize the circuit $\mathcal{O}^2$ and call the result $\mathcal{L}\mathcal{R}$, where $\mathcal{L}$ is the maximal prefix of $\mathcal{O}$ in the optimization of $\mathcal{O}^2$.  We then optimize $\mathcal{O}^3$.  Provided optimizations only occur near the boundaries between blocks, we can remove the prefix $\mathcal{L}$ and the suffix $\mathcal{R}$ from the optimized version of $\mathcal{O}^3$, and call the remaining circuit $\mathcal{C}$.  Assuming we can find such $\mathcal{L}$, $\mathcal{C}$, and $\mathcal{R}$ (which is always the case in practice), then we can simplify $\mathcal{O}^t$ to ${\mathcal L}{\mathcal C}^{t-2}{\mathcal R}$.

\item\emph{Toffoli decomposition:}
Many quantum algorithms are naturally described using Toffoli gates.  Our optimizer can handle Toffoli gates with both positive and negative controls.  Since we ultimately aim to express circuits over the gate set $\{\notgate,\cnotgate,\hgate,\rzgate\}$, we must decompose the Toffoli gate in terms of these elementary gates.  We take advantage of different ways of doing this to improve the quality of optimization.

Specifically, we expand the Toffoli gates in terms of one- and two-qubit gates using the identities shown in \fig{toffoli}, keeping in mind that we also obtain the desired Toffoli gate by exchanging $\tgate$ and $\tgate^{\dagger}$ in those circuit decompositions (because the Toffoli gate is self-inverse).  Initially, the optimizer leaves the \emph{polarity} of $\tgate/\tgate^\dagger$ gates (i.e., the choice of which gates include the dagger and which do not) in each Toffoli decomposition undetermined.  The optimizer symbolically processes the indeterminate $\tgate$ and $\tgate^{\dagger}$ gates by simply moving their locations in a given quantum circuit, keeping track of their relative polarities.  The optimization is considered complete when movements of the indeterminate $\tgate$ and $\tgate^{\dagger}$ gates cannot further reduce the gate count. Finally, we choose the polarities of each Toffoli gate (subject to the fixed relationships between them) with the goal of minimizing the $\tgate$ count in the optimized circuit.  We perform this minimization in a greedy way, choosing polarities for each Toffoli gate in the order of appearance of the associated $\tgate/\tgate^\dagger$ gates in the nearly-optimized circuit, so as to reduce the $\tgate$ count as much as possible.

Overall, this polarity selection process takes time $O(g)$.  After choosing the polarities, we run \routine{Rdouble} and \routine{Rsingle}, since particular choices of polarities may lead to further cancellations of the $\cnotgate$ gates and single-qubit gates that were otherwise not possible due to the presence of the indeterminate gates blocking the desired commutations.
\end{itemize}





\section*{Acknowledgements}

This work was supported in part by the Army Research Office (grant W911NF-16-1-0349), the Canadian Institute for Advanced Research, and the National Science Foundation (grant CCF-1526380).

This material was partially based on work supported by the National Science Foundation during DM's assignment at the Foundation. Any opinion, finding, and conclusions or recommendations expressed in this material are those of the authors and do not necessarily reflect the views of the National Science Foundation.







\end{document}